\documentclass[]{spie}  

\usepackage{amsmath,amsfonts,amssymb}
\usepackage{graphicx}
\usepackage[colorlinks=true, allcolors=blue]{hyperref}

\title{Topological Designs for Scalar Vortex Coronagraphs
}

\author[a]{Niyati Desai}
\author[a]{Jorge Llop-Sayson}
\author[a]{Arielle Bertrou-Cantou} 
\author[b]{Garreth Ruane}
\author[b]{A J Eldorado Riggs}
\author[b]{Eugene Serabyn}
\author[a]{Dimitri Mawet}
\affil[a]{Department of Astronomy, California Institute of Technology, 1200 East California Blvd., Pasadena, CA, 91125}
\affil[b]{Jet Propulsion Laboratory, California Institute of Technology, Pasadena, CA 91109, USA}

\authorinfo{Send correspondence to ndesai2@caltech.edu}

\begin{document} 
\maketitle

\begin{abstract}

The detection and characterization of Earth-like exoplanets around Sun-like stars for future flagship missions requires coronagraphs to achieve contrasts on the order of $10^{-10}$ at close angular separations and over large spectral bandwidths ($\geq$20\%).  We present our progress thus far on exploring the potential for scalar vortex coronagraphs (SVCs) in direct exoplanet imaging. SVCs are an attractive alternative to vector vortex coronagraphs (VVCs), which have recently demonstrated $6\times10^{-9}$ raw contrast in $20\%$ broadband light but are polarization dependent. SVCs imprint the same phase ramp on the incoming light and do not require polarization splitting, but are inherently limited by their chromatic behavior. Several SVC designs have been proposed in recent years to solve this issue by modulating or wrapping the azimuthal phase function according to specific patterns. For one such design, the staircase SVC, we present our best experimental SVC results  demonstrating raw contrast of $2\times10^{-7}$ in $10\%$ broadband light. Since SVC broadband performance and aberration sensitivities are highly dependent on topology, we conducted a comparative study of several SVC designs to optimize for high contrast across a range of bandwidths. Furthermore, we present a new coronagraph optimization tool to predict performance in order to find an achromatic solution.

\end{abstract}


\section{INTRODUCTION}
\label{sec:intro}  

Of the over 5,000 exoplanets that have been detected, only about a dozen have ever been directly imaged.\cite{Akenson2013} Techniques such as transit and radial velocity make up the majority of exoplanet detections that are confirmed today; however, these methods are not sensitive enough to allow for the study of key molecules in exoplanet atmospheres or surface temperatures over a wide range of planet types and separations from the host star. In particular, direct imaging is the only viable method for the detection and characterization of long-period exoplanets on solar system scales and in particular Earth-like planets orbiting Sun-like stars. The Astro2020 Decadal Survey has recognized direct imaging of exoplanets as a valuable research direction for space telescope mission concepts such as the Habitable Exoplanet Explorer (HabEx) and the Large Ultra-violet,
Optical, Infrared (LUVOIR) Surveyor\cite{Astro2020_Report,HabExReport, LUVOIRReport}.

Direct imaging is very challenging because Earth-like exoplanets are on the order of $10^{10}$ times fainter than their host star in visible and infrared. Furthermore, for the purpose of atmospheric spectroscopy it is necessary to be able to directly image the target over a large spectral band. For these reasons, the selection of a coronagraph depends on its ability to suppress starlight, but also factors like high off-axis throughput (e.g., from a planet), a small inner working angle, and insensitivity to tip/tilt errors and other low-order aberrations. 

Over the last fifteen years, the vortex coronagraph has been explored as a viable alternative to the classic Lyot coronagraph for high contrast imaging because it offers a desirable balance between these factors \cite{Swartzlander2005, Mawet2005, Foo2005, Lee2006, Swartzlander2008}. Vortex coronagraphs come in two flavors: vector vortex coronagraphs (VVCs) and scalar vortex coronagraphs (SVCs). Each offers its own advantages and conversely unique limitations. The primary difference lies in how vector and scalar vortex focal plane masks differently imprint an optical vortex onto incoming wavefronts. 

The development of VVCs has received much more attention than SVCs and they have recently been successfully implemented in several ground based observatories. On-sky demonstration and operation of VVCs at Palomar Observatory\cite{Mawet2009}, Keck Observatory\cite{Wang2020,Serabyn2017}, and the VLT\cite{Wagner2021} have proven the capabilities of vortex coronagraphs while experimental testing and further development at Caltech and JPL continues to push their potential\cite{Serabyn2019,Ruane2020,Ruane2022}.

The earliest literature on SVCs only dates back to Swartzlander et al.~2006\cite{Swartzlander2006} who explored their potential and suggested possible limitations. Later, Swartzlander et al.~2008\cite{Swartzlander2008} reported the first demonstration of an SVC. More recent literature by Ruane et al.~2019\cite{Ruane2019} proposes the theoretical and simulated potential that SVCs offer. Finally, Galicher et al.~2020\cite{Galicher2020} and Desai et al.~2021\cite{Desai2021} provide preliminary experimental results from SVCs and mention the achromatic limitations discussed in Section~\ref{subsec:Advs}.

\section{MOTIVATION}
\label{sec:motivation}  

\subsection{Scalar Vortex Advantages and Challenges}
\label{subsec:Advs}

A vortex coronagraph diffracts the starlight that is centered on the mask away from the center to be blocked by a Lyot stop downstream in the pupil plane\cite{Swartzlander2005, Mawet2005, Foo2005, Lee2006, Swartzlander2008}. As mentioned in Section~\ref{sec:intro}, there are two flavors of vortex coronagraph: scalar and vector. SVCs and VVCs both imprint a phase ramp onto an incoming wavefront to create an optical vortex, but they differ in the mechanism each uses to do so.

Scalar vortex masks are manufactured by varying the substrate thickness or index of refraction and they fundamentally interact with light differently than vector vortex masks. SVCs turn incoming wavefronts into an optical vortex through longitudinal phase delays, whereas VVCs use geometric phase shifts. Vector vortex masks are essentially a half wave plate with a spatially varying fast axis, which results in a polarization dependence. They are often manufactured from liquid crystal polymers~\cite{Mawet2009LiquidCrystal,Ganic2002}, subwavelength gratings~\cite{Mawet2005Subwavelength}, photonic crystals~\cite{Murakami2013}, or metamaterials~\cite{Genevet2012}.

A scalar vortex mask's primary limitation comes from the fact that a longitudinal phase delay is inherently chromatic in behavior. Equation \ref{eq:transmission} shows the complex transmission function of an SVC and includes the wavelength dependence\cite{Ruane2019}. Here the desired phase ramp to be imprinted is $\exp(i l \theta)$ where $l$ is the vortex charge, or the number of phase wraps the light does in one wavelength. An ideal vortex coronagraph requires $l$ to be a nonzero, even integer value for the starlight to be completely diffracted outside of the Lyot stop~\cite{Swartzlander2005, Mawet2005, Foo2005, Lee2006}.

\begin{equation}
\label{eq:transmission}
    t=\exp \left(i \frac{l \lambda_{0}}{\lambda} \theta\right)
\end{equation}

At the design wavelength $\lambda_{0}$, an SVC creates a perfect optical vortex with a spiral phase ramp described by $\exp(i l \theta)$. However as the wavelength is offset from the central design wavelength, the charge $l$ becomes non-integral for $\lambda \ne \lambda_{0}$ and the SVC's ability to perform light suppression decreases--hence it is chromatic.

A VVC with charge $l$ still effectively imprints a spiral phase ramp  onto the incoming light, but creates two different ramps $\exp(\pm i l \theta)$ depending on the polarization states of the light. Equation \ref{eq:jones} shows the Jones matrix in the circular polarization basis that demonstrates how the VVC focal plane mask modifies the wavefront differently based on the polarization state of the light~\cite{Ruane2019, Mawet2010}. Additionally there is some amount of polarized stellar leakage $c_{L}$ due to imperfect retardance. 

\begin{equation}
\label{eq:jones}
\mathbf{M}_{\circlearrowright}=c_{V}\left[\begin{array}{cc}
0 & e^{i l \theta} \\
e^{-i l \theta} & 0
\end{array}\right]+c_{L}\left[\begin{array}{cc}
1 & 0 \\
0 & 1
\end{array}\right]
\end{equation}

In the Jones matrix, note the sign of the imprinted phase ramp is opposite for right and left circularly polarized light. VVC behavior is fundamentally limited by this polarization dependence because, in order to do wavefront sensing and control, one phase ramp must be isolated. This is a major drawback because half the throughput is eliminated by introducing an analyzer and polarizer to filter only one polarization state. 

This trade-off between polarization dependence and chromaticity is the primary consideration between vector and scalar vortex coronagraphs. A further discussion of the VVC's polarization dependence limitation is detailed in Ruane et al. 2019\cite{Ruane2019}.

\subsection{Addressing the Chromaticity}
\label{subsec:Achromatizing}

Although SVCs and VVCs fundamentally share the same starlight suppression goal, in the last fifteen years, VVCs have been more researched and pushed to their light suppression limits. They have recently been demonstrated to reach raw contrast of 2e-9 and 6e-9 in $10\%$ and $20\%$ optical bandwidths, respectively~\cite{Ruane2022}. However, their polarization dependent limitations motivate research in alternate avenues, namely scalar vortex coronagraphs and addressing their chromatic behavior. 

Since scalar vortex coronagraphs have not received as much attention, only a few studies in the literature have outlined their potential for high contrast imaging and directly addressed their chromatic limitations\cite{Swartzlander2006,Ruane2019,Galicher2020,Desai2021}. Several different designs were proposed in recent years by Ruane et al.~\cite{Ruane2019} and Galicher et al.~\cite{Galicher2020} to address the chromatic behavior of SVCs by modulating or wrapping the azimuthal phase function according to specific patterns.

\begin{figure} [ht]
\begin{center}
\begin{tabular}{c} 
\includegraphics[height=6cm]{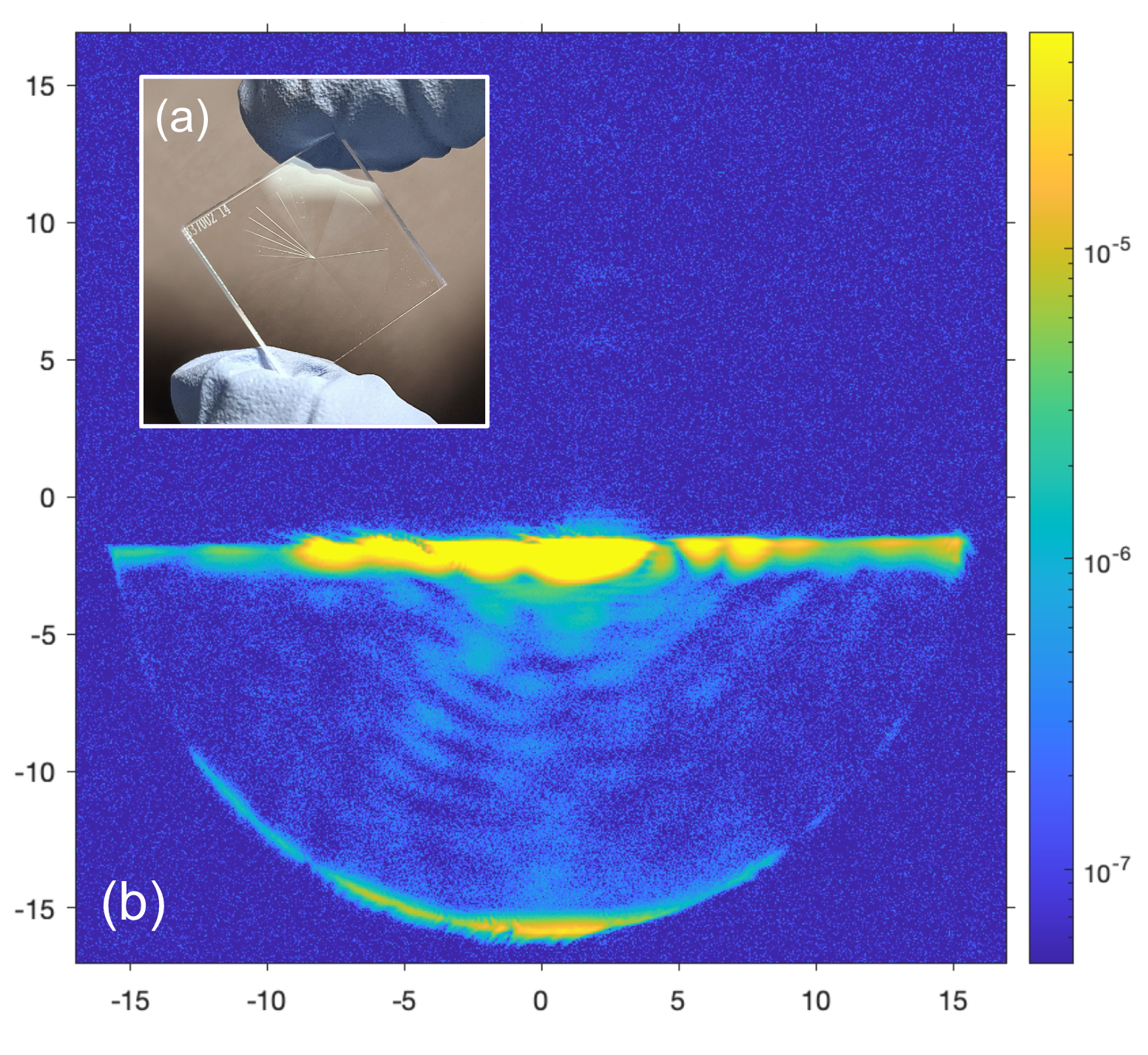}
\end{tabular}
\end{center}
\caption[fig:darkhole] 
{ \label{fig:darkhole} 
(a) A charge 6 staircase SVC mask fabricated in fused silica (see Desai et al.~\cite{Desai2021} for further details on this mask). (b) The final dark hole result of this staircase mask with electric field conjugation on the In Air Coronagraph Testbed (IACT) at JPL. It is a 10\% bandwidth dark hole from 3-10 $\lambda$/D with an average contrast of 2.2$\times10^{-7}$.}
\end{figure}

We have designed, fabricated, and tested one such design (staircase vortex) on the High Contrast Spectroscopy Testbed (HCST) at Caltech and on the In Air Coronagraph Testbed (IACT) at JPL. The phase mapping and phase profile for this design can be found in Section~\ref{sec:topo}, and further details on this mask can be found in Desai et al.~\cite{Desai2021}. Our best experimental result with this charge 6 staircase SVC is an average contrast of 2.2e-7 in $10\%$ bandwidth with three sub-bandpasses.  We performed estimation and control for a single DM using the electric field conjugation (EFC) algorithm\cite{Give'On2009} to dig a 180$^\circ$ dark hole from 3-10 $\lambda$/D shown in Figure~\ref{fig:darkhole}.

\begin{figure} [ht]
\begin{center}
\begin{tabular}{c} 
\includegraphics[height=4.6cm]{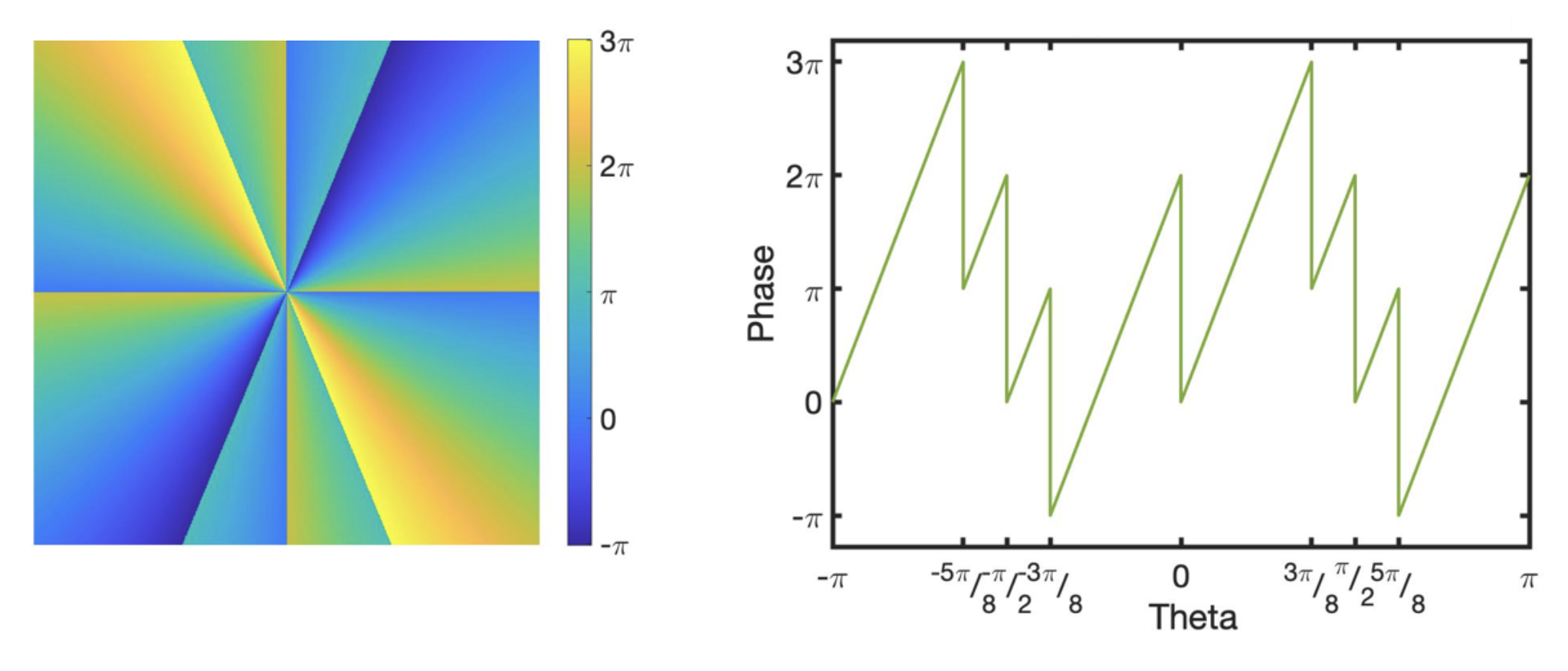}
\end{tabular}
\end{center}
\caption[fig:galicher] 
{ \label{fig:galicher} 
Phase mapping (left) and azimuthal phase profile (right) for the charge 8 wrapped vortex focal plane mask proposed by Galicher et al.~2020. The same wrapping technique was applied to create the charge 6 wrapped vortex in Figure~\ref{fig:phasemaps}(d). and Figure~\ref{fig:phaseprofs}(d).}
\end{figure}

Galicher et al.~2019 demonstrated that cleverly wrapping the phase function resulted in improved broadband performance~\cite{Galicher2020}. The phase mapping and phase profile of the charge 8 `Galicher Vortex' in Fig.~\ref{fig:galicher} show the unique phase wrapping proposed. Requirements were placed on phase discontinuities to be jumps in multiples of 2$\pi$ and overall the resulting phase ramp had to be that of a charge 8 vortex. Because this design was reported to improve broadband performance, one of the motivations for this study was to more generally understand phase wrapping comparatively.

To investigate how the topology of a scalar vortex mask affects its performance, and specifically its chromaticity, we performed a direct comparison of several different topologies and present our findings here.

\section{Comparing Scalar Vortex Topologies}
\label{sec:topo}

\begin{figure} [ht]
\begin{center}
\begin{tabular}{c} 
\includegraphics[height=8cm]{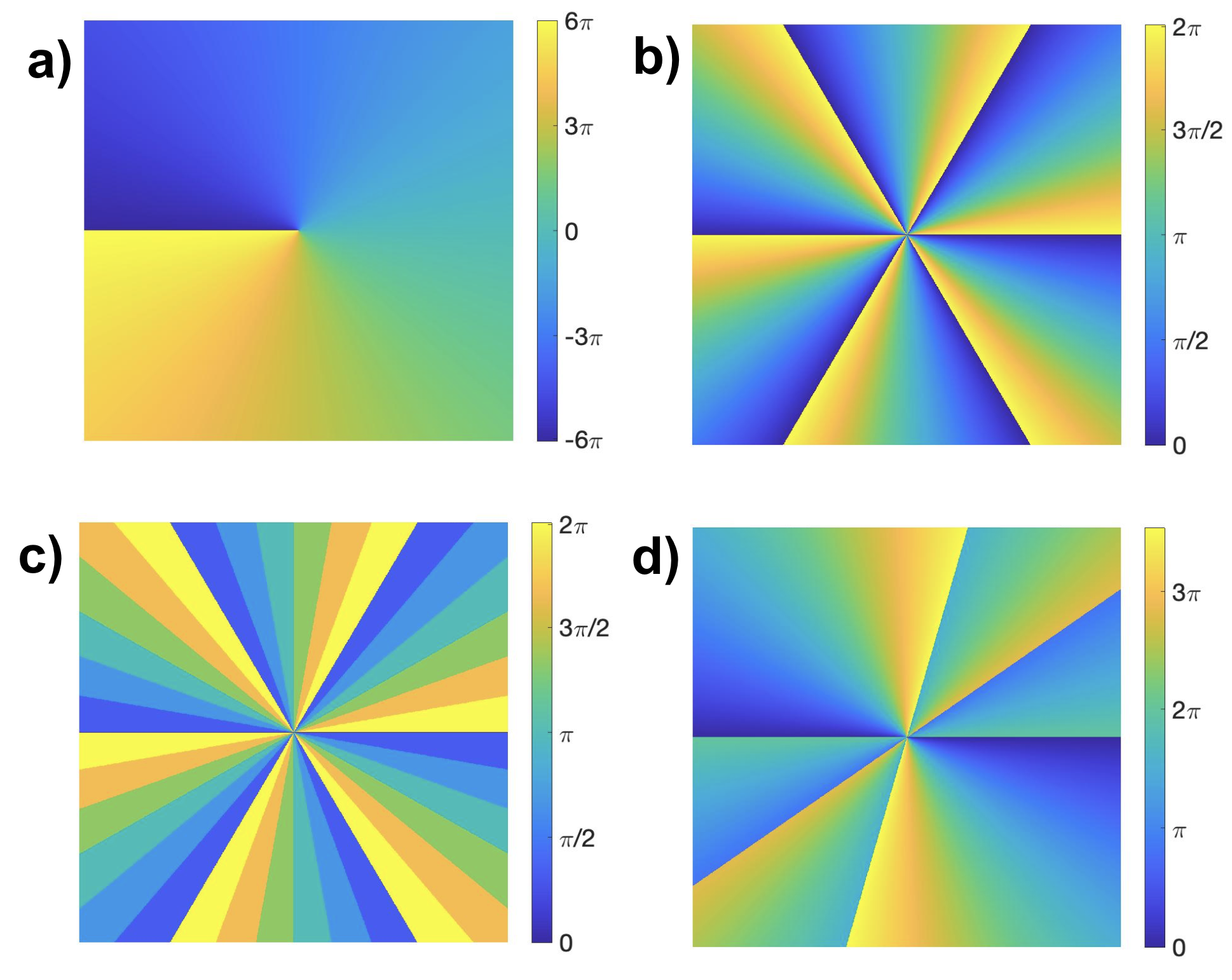}
\end{tabular}
\end{center}
\caption[fig:phasemaps] 
{ \label{fig:phasemaps} 
Phase mappings for four charge 6 scalar vortex focal plane mask topologies: (a) classic vortex, (b) sawtooth vortex, (c) staircase vortex, (d) wrapped vortex. The staircase vortex (c) was originally proposed in Ruane et al.~2019 and the wrapped vortex (d) was inspired by the wrapping proposed by Galicher et al.~2020. }
\end{figure}

\begin{figure} [ht]
\begin{center}
\includegraphics[height=10cm]{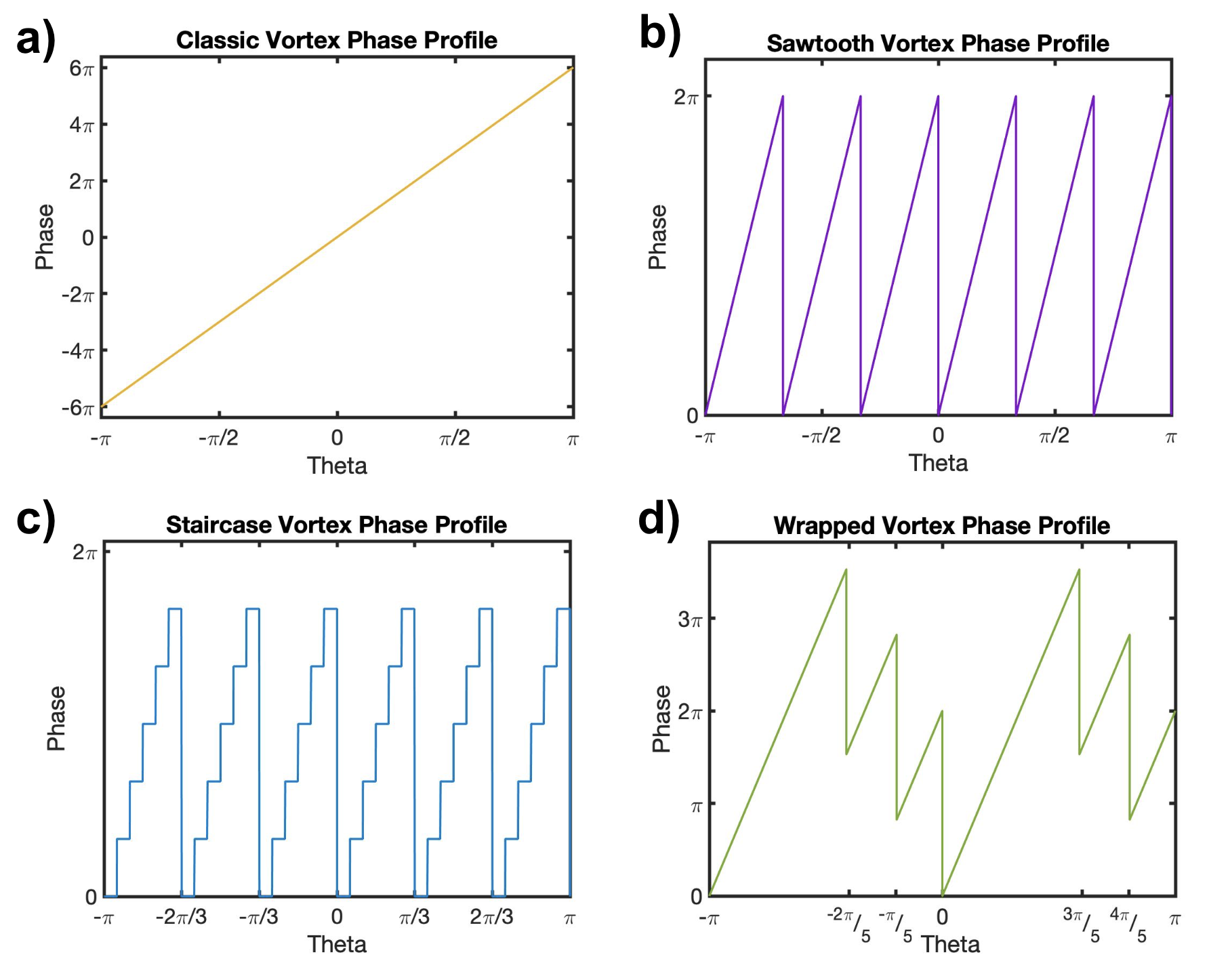}
\end{center}
\caption[fig:phaseprofs] 
{ \label{fig:phaseprofs} 
Phase profiles for four charge 6 scalar vortex focal plane mask topologies: (a) classic vortex, (b) sawtooth vortex, (c) staircase vortex, (d) wrapped vortex corresponding to the phase mappings shown in Figure~\ref{fig:phasemaps}.}
\end{figure}

We simulated several different SVC topologies of charge 6 and charge 8 to perform a direct comparison of their performance. Figure~\ref{fig:phasemaps} shows the phase mappings only of a charge 6 classic vortex (a), sawtooth vortex (b), staircase vortex (c), and wrapped vortex (d). Figure~\ref{fig:phaseprofs} shows the corresponding azimuthal profiles of each of the phase mappings shown in Figure~\ref{fig:phasemaps}.

\textbf{Classic Vortex:}
The first topology of SVC we considered is the classic vortex. A classic vortex SVC is a simple phase ramp wrapping around the optical axis, and its slope is proportional to the charge of the vortex. For example, in Figure~\ref{fig:phasemaps}, (a) is a charge 6 classic vortex, so its phase is a linear ramp from $-6\pi$ to $+6\pi$. For a charge 8 classic vortex, the phase ramp goes from $-8\pi$ to $+8\pi$.

\textbf{Sawtooth Vortex:}
We call the second SVC topology considered in this study the sawtooth vortex SVC because in Figure~\ref{fig:phaseprofs} the profile of this mask (b) looks like a sawtooth pattern. In the sawtooth vortex design, $n$ phase ramps range from 0 to $2\pi$ and the charge of the vortex dictates the number of phase ramps, and therefore the slope. For example in (b) of Figures~\ref{fig:phasemaps} and~\ref{fig:phaseprofs}  the charge 6 sawtooth vortex has 6 ramps, each of which starts at a phase of 0 and goes to $2\pi$. This is the design for the fabricated device in Figure~\ref{fig:darkhole} which was tested on IACT.

\textbf{Staircase Vortex: }
The third topology modeled in this study is the staircase vortex design from Ruane et al.~2019\cite{Ruane2019}, whose name is apparent from its phase profile shown in Figure~\ref{fig:phaseprofs}. Instead of several continuous linear phase ramps, this design allows for several sectors with flat steps at heights ranging from 0 to $2\pi$ within each. For example, Figure~\ref{fig:phasemaps} (c) shows a charge 6 staircase vortex with 6 sectors and 6 steps in each sector.

\textbf{Wrapped Vortex: }
The fourth topology is inspired by the phase wrapping method proposed by Galicher et al. in 2019. As mentioned in Section~\ref{sec:motivation}, that study demonstrated that cleverly wrapping the phase function resulted in improved broadband performance for a charge 8 SVC. We applied a similar wrapping principle to create a charge 6 vortex and Figures~\ref{fig:phasemaps} and ~\ref{fig:phaseprofs} (d) show that design. This profile was obtained with a Markov chain Monte Carlo (MCMC) optimization of the starlight suppression, using the azimuthal position of the $2\pi$ jumps as parameters.


\section{Simulations}
\label{sec:sims}
We used the Fast Linearized Coronagraphic Optimizer (FALCO)\footnote{\url{https://github.com/ajeldorado/falco-matlab}} software package for simulating the point spread function (PSF) of a star, and wavefront propagation through all subsequent pupils and optics including our focal plane mask~\cite{Riggs2018}. To simulate the various focal plane masks proposed in section~\ref{sec:topo}, we provided high resolution phase mappings like the ones shown in Figure~\ref{fig:phasemaps} and adjusted resolution parameters in FALCO's wavefront propagation to find an appropriate simulation environment for this study. 

\subsection{Parameters of FALCO Simulation}
\label{sec:params}

We created a new focal plane mask model for each SVC topology (classic, sawtooth, staircase, wrapped) for charge 6 and for charge 8 and kept the other simulation parameters of the testbed constant. In FALCO there are three types of optical models for each coronagraph: full, compact, and Jacobian. The full model represents the true physics of a testbed, whereas the compact model represents the best knowledge of the setup that is available to the estimator and controller. For this reason, the full model is used only in simulations rather than actual testbed operations. 

We found focal plane phase mask resolution parameters of 8 and 4 pixels per $\lambda$/D for the full and compact models, respectively, yielded meaningful contrasts which were not limited by simulation resolutions. Furthermore, a new Tukey window function was recently added to FALCO which does small field of view (FOV) fine resolution FFT within some radius and full FOV course resolution DFT outside it. We also adjusted these parameters and found an inner radius of 10 $\lambda$/D and an outer radius of 17 $\lambda$/D covered the entire dark hole over 3-10 $\lambda$/D and proved to be sufficient resolutions to carry out these simulations within a reasonable computation time.

We perform simulations in monochromatic and in polychromatic light spanning 1\% to 20\% bandwidth with 9 sub-bandpasses for EFC\cite{Give'On2009}. We simulate digging a dark hole with 15 iterations of EFC between 3-10 $\lambda$/D for a 360$^\circ$ dark hole and two $34 \times 34$ actuator deformable mirrors, performed with the same FALCO operations that we typically employ in our testbed environment.



\subsection{2D Simulation Results}
\label{sec:results}

One of our primary metrics for measuring the performance of these SVC topologies was the final raw contrast at the focal plane. Raw contrast is defined as the averaged intensity of the simulated coronagraphic image at the focal plane divided by the peak intensity of the unocculted pseudo-star PSF image.

\begin{figure} [ht]
\begin{center}
\includegraphics[height=6cm]{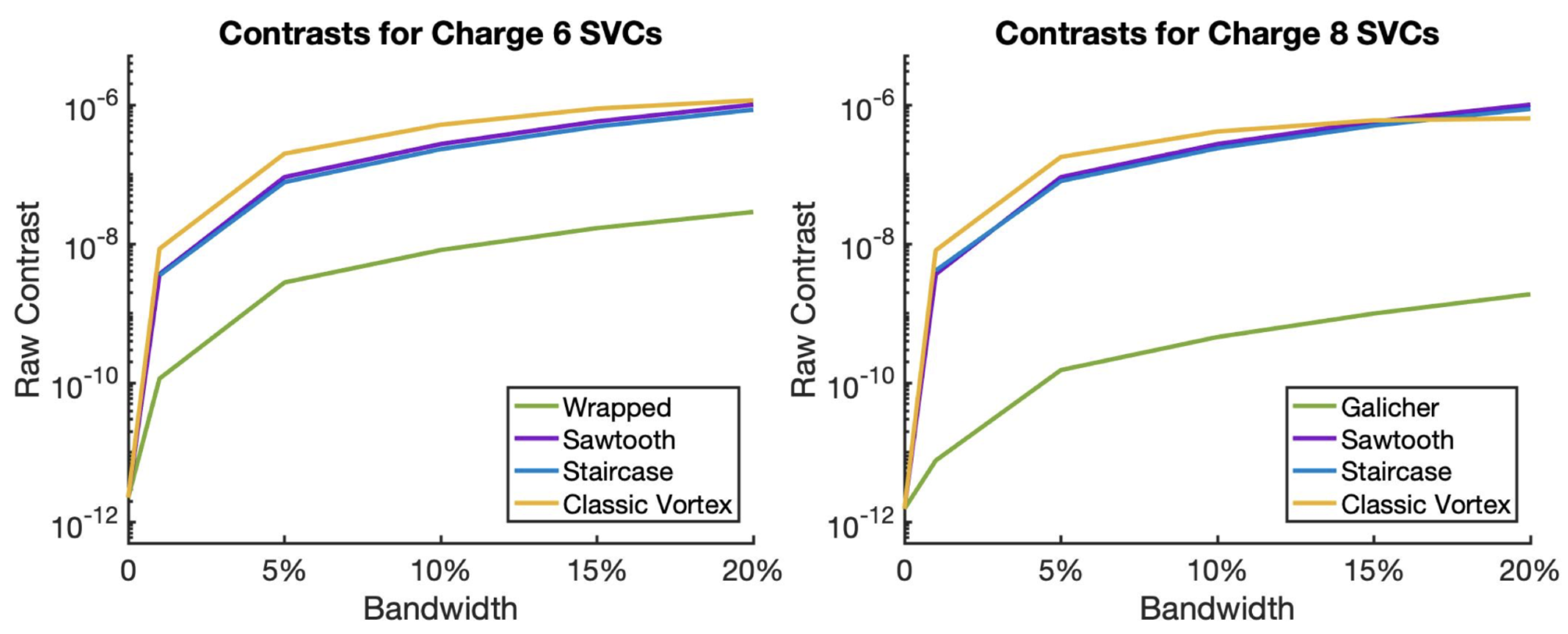}
\end{center}
\caption[fig:contrasts] 
{ \label{fig:contrasts} 
Raw contrasts for charge 6 SVCs (left) and charge 8 SVCs (right) simulated using FALCO packages.}
\end{figure}

The contrast curves in Figure~\ref{fig:contrasts} show the starlight suppression performance as a function of bandwidth of the four topologies of SVCs (with corresponding colors to the profiles in Figure~\ref{fig:phaseprofs}).  The left figure shows a comparison between four SVC topologies in a charge 6 mask and the right shows a comparison between the same four SVC topologies in a charge 8 mask.

These contrast curves display how chromatic each topology is. For monochromatic light, all four topologies achieve raw contrasts of around $10^{-12}$, likely limited by numerical noise. However as the bandwidth increases, the contrasts for the sawtooth, staircase, and classic vortex quickly deteriorate to $10^{-8}$ for $1\%$ bandwidth and to around $10^{-7}$ for $5\%$ bandwidth. The green contrast curves in both plots of Figure~\ref{fig:contrasts} show significantly better broadband performance than the three other topologies and the contrasts for the charge 8 Galicher mask agree with those reported in the original study~\cite{Galicher2020}. We were able to confirm that wrapping the phase ramp improves broadband performance by also presenting our charge 6 wrapped mask contrast results in the left plot in Figure~\ref{fig:contrasts}. As expected, the charge 6 wrapped mask contrasts did not outperform the charge 8 Galicher mask, however the new charge 6 wrapped design still offers a two order of magnitude improvement at broadband over the other charge 6 topologies.

\section{Modal Decomposition}
\label{sec:modaldecomp}
Originally for the pursuit of understanding the comparative performance of these topologies better, Ruane et al. 2019 provides an analysis tool depending only on the phase profile of a topology. We developed and used this tool to get a more extensive understanding of SVC behaviors. We present this tool for predicting the performance of any azimuthally varying SVC topology. It is important to understand why various topologies yield different contrast curves and this modal decomposition tool provides a possible insight in explaining this.

\subsection{Modal Decomposition for SVC performance analysis}

Modal decomposition is a way to break down and visualize the power distribution of an SVC, which gives insight into stellar leakage and sensitivity to low order aberrations, as well as the effective topological charge for any potentially complex phase profile. Since an optical vortex is a Fourier series, in the exponential form, each coefficient can be linearly decomposed as shown below.

Any optical vortex can be written as:
\begin{equation}
\label{eq:fourier}
    t(\theta,\lambda)=\Sigma_{m} C_{m}(\lambda)e^{im\theta}
\end{equation}

with

\begin{equation}
\label{eq:coeffs}
    C_{m}(\lambda) = \frac{1}{2\pi} \int_{\pi}^{-\pi} t(\theta,\lambda) e^{-im\theta} d\theta
\end{equation}

And each mode \textit{m} corresponds to a charge m vortex. This can be interpreted to predict the behavior of a new SVC topology as a combination of differently charged vortexes.

\begin{figure} [ht]
\begin{center}
\includegraphics[height=6cm]{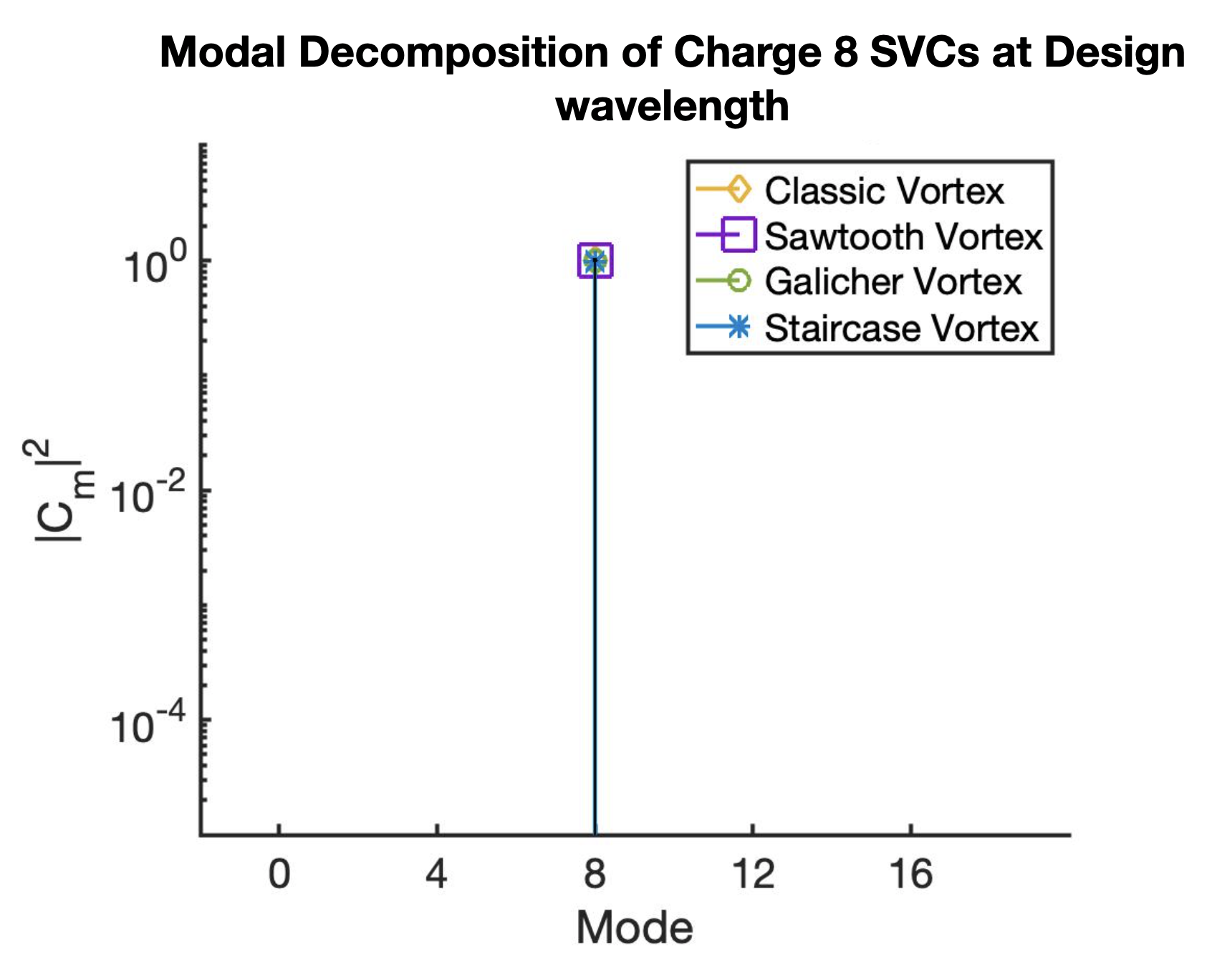}
\end{center}
\caption[fig:modalcentral] 
{ \label{fig:modalcentral} 
This modal decomposition shows four different charge 8 scalar vortex topologies at the design central wavelength. The overlapping peaks at mode 8 are expected since they all perform comparably in monochromatic light.}
\end{figure}

We applied this technique to the SVC topologies mentioned previously to compare their predicted performances. As shown in Figure~\ref{fig:modalcentral}, at the design central wavelength, only the expected mode is found and there is consistent behavior across all the SVC topologies. In this figure, several charge 8 topologies are considered so the peaks all appear at mode 8. This is expected since all the SVCs should imprint the same desired phase ramp on the light at the design central wavelength.

\begin{figure} [h]
\begin{center}
\includegraphics[height=5.5cm]{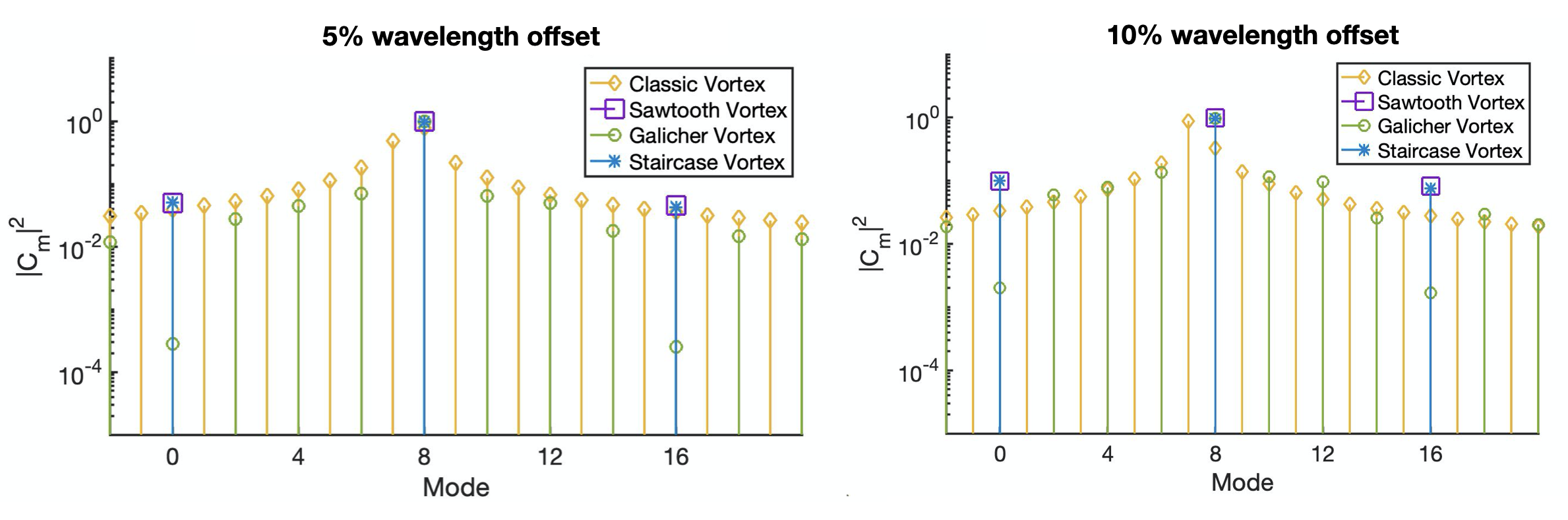}
\end{center}
\caption[fig:modaloffset] 
{ \label{fig:modaloffset} 
These two modal decompositions for charge 8 SVCs at a $5\%$ wavelength offset (left) and a $10\%$ wavelength offset (right) show how each of the classic vortex, sawtooth vortex, Galicher vortex, staircase vortex topologies differ in behavior for broadband light.}
\end{figure} 

However when a modal decomposition is performed at a wavelength offset from the central wavelength, the chromatic behavior of each SVC is revealed. In Figure~\ref{fig:modaloffset} at a $5\%$ offset (left), mode 8 is still the primary peak, however a peak emerges at mode zero which corresponds to zeroth order leakage. Zeroth order leakage corresponds to starlight passing straight through the coronagraph without being affected by the vortex at all and therefore not being suppressed. As seen in this figure, the Galicher wrapped vortex (green) has a significantly lower zeroeth order leakage, which explains the two orders of magnitude improvement in contrast at larger bandwidths seen in Figure~\ref{fig:contrasts}.

Furthermore, the peaks at even modes correspond to energy distributed to evenly charged vortex behaviors, whose behaviors are well studied. Odd modes, on the other hand, correspond to odd charged vortexes, where starlight suppression is deficient. This explains why the classic vortex plot (orange), which displays odds modes in the modal decomposition, exhibits the worst contrast performance in Figure~\ref{fig:contrasts}. 

We believe the peaks corresponding to low even charge vortexes could be correlated to low order wavefront aberrations known to more negatively affect vortex coronagraphs with low charges~\cite{Ruane2020}. Peaks at even modes larger than the desired charge are not expected to be as problematic since higher charge vortexes do not have high sensitivities to aberrations, although they do affect the throughput and inner working angle.

\subsection{Modal Decomposition Potential for Optimizing New Designs}
The modal decomposition representation of a phase mask can also be used to reverse-engineer a new mask design. This is done by selecting certain peaks to suppress: for instance the zeroth mode, odd modes, or small even modes, and then mitigating their modal intensity. Then an inverse Fourier transform is applied on this distribution to result in a new phase profile. 

We find this optimization to be challenging given our modest understanding of the interplay between modes combined with the difficulties associated with optimizing for a large bandwidth. When trying to suppress one mode, others change in intensity in a way we have yet to understand. Furthermore, covering a large spectral band and nulling the zeroth order at a given wavelength increases the intensity of other modes in other wavelengths, thus deteriorating low-order aberrations sensitivity and starlight suppression of the mask. 

\begin{figure} [ht]
\begin{center}
\includegraphics[height=5.5cm]{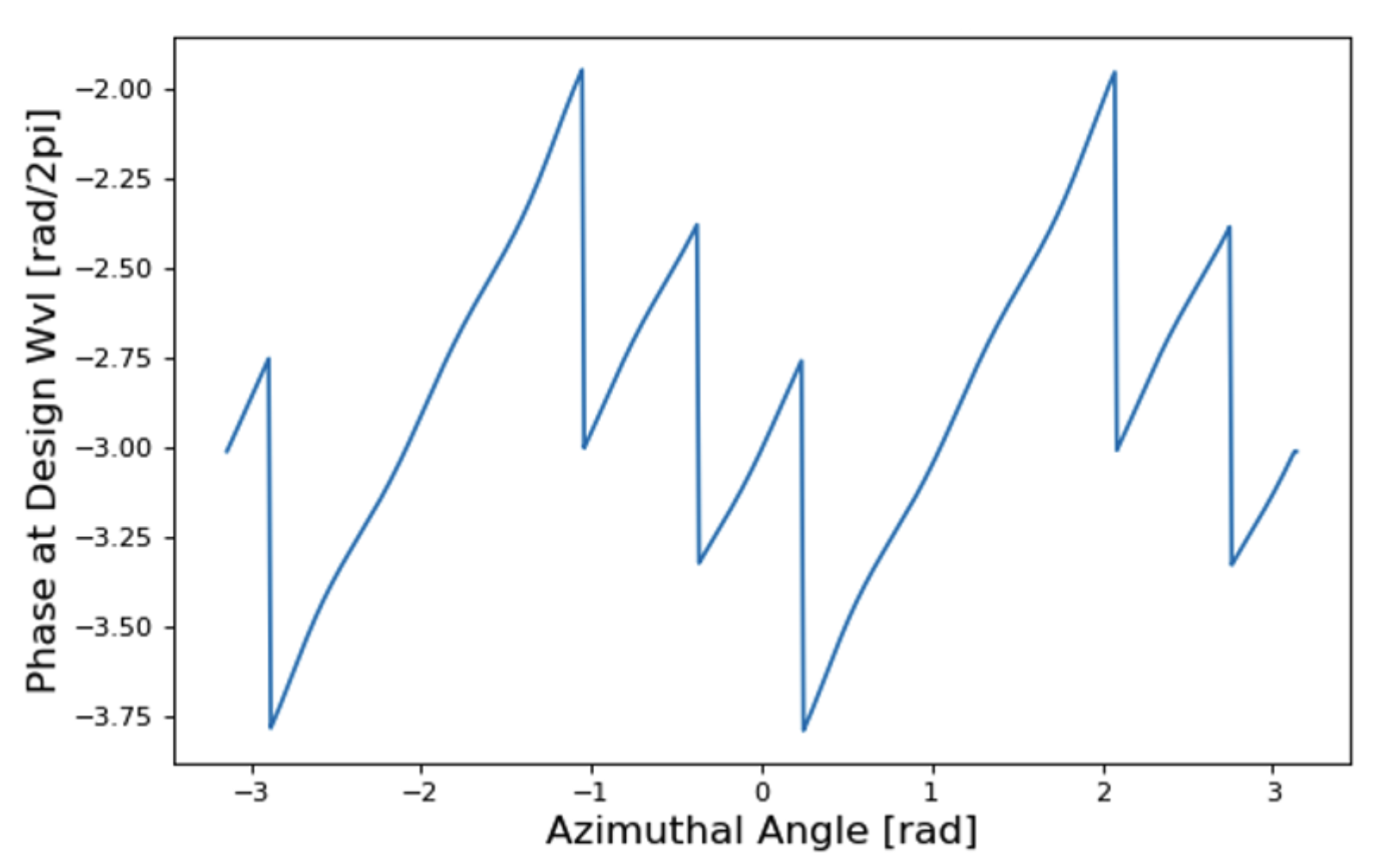}
\end{center}
\caption[fig:whackamolephase] 
{ \label{fig:whackamolephase} 
 A resulting phase profile from a preliminary attempt at optimization using modal decomposition to suppress the peak at the zeroth mode. The subtle deviations from a constant phase ramp cause the partial suppression of the zeroth mode, as seen in Figure~\ref{fig:whackamoledecomp}.}
\end{figure}

\begin{figure} [ht]
\begin{center}
\includegraphics[height=5cm]{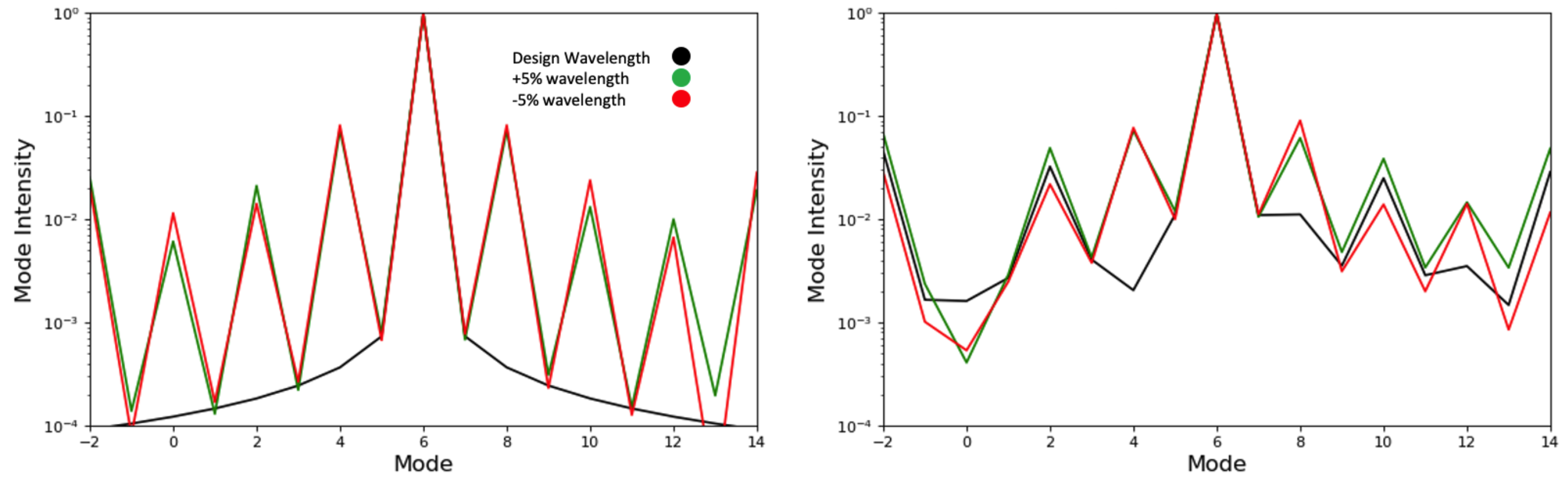}
\end{center}
\caption[fig:whackamoledecomp] 
{ \label{fig:whackamoledecomp} 
The left plot is the original modal decomposition for a charge 6 Galicher vortex at the design wavelength (black), at $+5\%$ wavelength offset (green), and at a $-5\%$ wavelength offset (red). The right plot is the the modal decomposition after nulling the zeroth order. Although the zeroth mode is partially suppressed, other unwanted modes have an increased intensity.}
\end{figure}

Figure~\ref{fig:whackamolephase} shows an optimized phase profile that resulted when carrying out this procedure and Figure~\ref{fig:whackamoledecomp} shows the corresponding modal decompositions before and after suppressing the peak at mode 0. This optimization process is in early stages of development and for this preliminary attempt we use scipiy's L-BFGS-B (limited memory Broyden–Fletcher–Goldfarb–Shanno) based algorithm to optimize the shape of the phase profile for a given number of resolution elements that describe the profile.

We present this optimized profile as a demonstration of how modal decomposition could be a useful tool for new topology design with SVCs. The phase profile in Figure~\ref{fig:whackamolephase} offers an improvement in terms of normalized intensity at 25\% bandwidth, with only a slight degradation of tip and tilt sensitivity, and no effect on planet throughput and inner working angle. More work on developing our tools is required to produce phase profiles that are unequivocally more optimal than the phase ramps presented here. Furthermore, manufacturing constraints, alignment tolerances, and Lyot stop considerations all have yet to be accounted for.

There remains much exploration space for development of new optimized mask designs like this one with the modal decomposition tool. One of the advantages is that modal decomposition will be able to clearly discern the effective charge for any complex mask design whose charge might not be obvious, like a sinusoidal or otherwise asymmetric topology. The optimized solution will be a balance between the desired contrast performance and an acceptable modal sensitivity.

\section{CONCLUSION}
\label{sec:conc}

We present a direct comparison of several scalar vortex phase mask topologies and confirm the benefits of phase wrapping: this technique improves the contrast performance by two or three orders of magnitude, for charge 6 and 8 respectively, with up to 20$\%$ broadband light. We developed a new tool for analyzing phase profile energy distributions in the form of modal decompositions of any azimuthally varying phase mask design. With this tool, we highlight that the gain in contrast provided by the wrapped vortexes is made at the cost of an increased sensitivity to low order aberrations compared to the sawtooth vortex. Furthermore we propose this as a possible optimization tool for the development of new mask designs and topologies. 

This study presents additional motivation to continue to explore scalar vortex mask development with new design fabrication and lab testing in the pursuit of improving scalar vortex coronagraph performance for exoplanet direct imaging.

\acknowledgments 
 This work was supported by the NASA ROSES APRA program, grant NM0018F610. Part of this research was carried out at the Jet Propulsion Laboratory, California Institute of Technology, under a contract with the National Aeronautics and Space Administration.

\bibliography{main} 
\bibliographystyle{spiebib} 

\end{document}